\documentclass{svproc}

\usepackage{url}

\usepackage{graphicx}
\usepackage{multicol}
\usepackage{footmisc}
\usepackage{xcolor}
\newcommand{\m}{\phantom{-}}
\newcommand{\I}{{\rm i}}

\begin{document}
\mainmatter              % start of a contribution
\title{The molecular nature of some $\Omega_c^0$ states}
\title{The molecular nature of some $\Omega_c^0$ states}  % abbreviated title (for running head)
\author{Gl\`oria~Monta\~na\inst{1}$^\dagger$ \and \`Angels~Ramos\inst{1} \and Albert~Feijoo\inst{2}}
\authorrunning{Gl\`oria~Monta\~na et al.} % abbreviated author list (for running head)
\tocauthor{Gl\`oria~Monta\~na, \`Angels~Ramos and Albert~Feijoo}

\institute{Departament de F\'{\i}sica Qu\`antica i Astrof\'{\i}sica and Institut de Ci\`encies del Cosmos (ICCUB), Universitat de Barcelona,
Mart\'{\i} i Franqu\`es 1, 08028 Barcelona, Spain, \\
\email{$^\dagger$gmontana@fqa.ub.edu}
\and
Nuclear Physics Institute, 25068 \v Re\v z, Czech Republic}

\maketitle              % typeset the title of the contribution

\begin{abstract}
A vector meson exchange model based on effective Lagrangians is used to build the meson--baryon interaction in the charm $+1$, strangeness $-2$ and isospin 0 sector. The s-wave scattering amplitudes resulting from the unitarization in coupled-channels show two resonances with masses and widths that are in very good agreement with those of the experimental $\Omega_c(3050)^0$ and $\Omega_c(3090)^0$ states observed by the LHCb collaboration.
The interpretation of these resonances as pseudoscalar meson--baryon molecules would mean the assignment $J^P=1/2^-$ to their spin--parity.
%\keywords{}
\end{abstract}
\section{Introduction}
A lot of theoretical effort in the field of baryon spectroscopy has lately arisen with the aim of explaining the inner structure of the five narrow $\Omega_c^0$ excited resonances observed by the LHCb Collaboration \cite{Aaij:2017nav} and possibly establishing their unknown values of spin--parity. Some works suggest a $css$ quark description within revisited quark models \cite{Agaev:2017jyt,Chen:2017sci,Karliner:2017kfm,Wang:2017hej,Wang:2017vnc,Cheng:2017ove,Wang:2017zjw,Chen:2017gnu}
while others propose a pentaquark interpretation \cite{Huang:2017dwn,An:2017lwg,Kim:2017jpx}.
Models that can describe some resonances as quasi-bound states of an interacting meson--baryon pair \cite{Hofmann:2005sw,JimenezTejero:2009vq,Romanets:2012hm} offer a complementary scenario, an approach that we have re-examined in \cite{Montana:2017kjw} in view of the new experimental data. 
It is plausible to expect that some excitations in the $C=1$, $S=-2$ sector can be obtained by adding a $u\bar{u}$ pair to the natural $css$ content of the $\Omega_c^0$, 
just as a pentaquark structure with a $c\bar{c}$ pair is more natural to explain the $P_c(4380)$ and $P_c(4450)$ excited nucleon resonances than an extremely high energy excitation of the three quark system. The hadronization of the five quarks can then lead to bound states, generated by the meson--baryon interaction in coupled channels.
This possibility is supported by the fact that the masses of the excited $\Omega_c^0$ baryons under study lie near the $\bar K\Xi_c$ and $\bar K\Xi_c^\prime$ thresholds and that they have been observed in the $K^-\Xi^+_c$ invariant mass spectrum.

\section{Formalism}
The sought resonances are dynamically generated as poles of the scattering amplitude $T_{ij}$, unitarized by means of the on-shell Bethe-Salpeter equation in coupled channels, which implements the resummation of loop diagrams to infinite order:
\begin{equation}\label{eq:BSeq}
T_{ij}=V_{ij}+V_{il}G_{l}T_{lj}\, .
\end{equation}

The $G_{l}$ function for the meson--baryon loop is regularized using the \textit{dimensional regularization} approach, which introduces the dependence on a subtraction constant $a_l(\mu)$ for each intermediate channel $l$ at a given regularization scale $\mu$ (see Eq.~(18) in \cite{Montana:2017kjw}).

The s-wave interaction kernel $V_{ij}$ is obtained from a t-channel vector meson exchange amplitude \cite{Hofmann:2005sw}, that has the same structure as the contact Weinberg-Tomozawa term in the $t\ll m_V$ limit:
\begin{equation}\label{eq:Vij}
 V_{ij}(\sqrt{s})=-C_{ij}\frac{1}{4f^2}\left(2\sqrt{s}-M_i-M_j\right) N_i N_j\, ,
\end{equation}
with $M_i$, $M_j$ and $E_i$, $E_j$ being the masses and the energies of the baryons, and $N_i$, $N_j$ the normalization factors ${N=\sqrt{(E+M)/2M}}$.

The coefficients $C_{ij}$ are obtained from the evaluation of the t-channel interaction diagram, with the effective Lagrangians of the hidden gauge formalism:
\begin{equation}\label{eq:vertexVPP}
\mathcal{L}_{VPP}=ig\langle\left[\partial_\mu\phi, \phi\right] V^\mu\rangle\,,
\end{equation}
\begin{equation}\label{eq:vertexBBV}
\mathcal{L}_{VBB}=\frac{g}{2}\sum_{i,j,k,l=1}^4\bar{B}_{ijk}\gamma^\mu\left(V_{\mu,l}^{k}B^{ijl}+2V_{\mu,l}^{j}B^{ilk}\right)\,,
\end{equation}
describing the vertices coupling the vector meson to pseudoscalars ($VPP$) and baryons ($VBB$), respectively, in the pseudoscalar meson--baryon ($PB$) scattering, and assuming $SU(4)$ symmetry \cite{Hofmann:2005sw}.

The interaction of vector mesons with baryons ($VB$) is built in a similar way and involves the three-vector $VVV$ vertex, which is obtained from:
\begin{equation}\label{eq:vertexVVV}
\mathcal{L}_{VVV}=ig\langle {\left[V^\mu,\partial_\nu V_\mu\right] V^\nu}\rangle \,.
\end{equation}
The resulting interaction is that of Eq.~(\ref{eq:Vij}) with the addition of the product of polarization vectors, $\vec{\epsilon}_i\cdot\vec{\epsilon}_j$.

The interaction potential is not $SU(4)$ symmetric even though this symmetry is encoded in the Lagrangians. It is broken with the use of the physical masses of the mesons and baryons involved, and a factor $\kappa_c=1/4$ that accounts for the higher mass of the charmed mesons exchanged in some of the non-diagonal transitions.  In fact, the transitions mediated by the exchange of light vector mesons like the dominant diagonal ones do not make explicit use of $SU(4)$ symmetry since they are effectively projected into their $SU(3)$ content.

The available $PB$ channels in the $(I,S,C)=(0,-2,1)$ sector are $\bar{K}\Xi_c (2964)$, $\bar{K}\Xi'_c (3070)$, $D\Xi (3189)$, $\eta \Omega_c (3246)$, $\eta' \Omega_c (3656)$, $\bar{D}_s \Omega_{cc} (5528)$, and $\eta_c \Omega_c (5678)$, with the corresponding thresholds in parenthesis. The doubly charmed $\bar{D}_s \Omega_{cc}$ and $\eta_c \Omega_c $ channels are neglected as their energy is much larger than that of the other channels. 
The matrix of $C_{ij}$ coefficients for the resulting 5-channel interaction is given in Table~\ref{tab:coeff}.

\begin{table}[h]
\caption{The $C_{ij}$ coefficients for the $(I,S,C)=(0,-2,1)$ sector of the  $PB$ interaction.}
\begin{center}
\begin{tabular}{l c c c c c}
\hline \\   [-3mm]
&{${\bar K}\Xi_c$}  & {${\bar K}\Xi_c^\prime$}  & { $D\Xi$}  & { $\eta\Omega_c^0$} &{$\eta^\prime\Omega_c^0$}  \\
\hline \\   [-3mm]
{${\bar K}\Xi_c$}         & $1$ & $0$ & $\sqrt{3/2}~\kappa_c$  & $\m\m 0$ & $\m\m 0$     \\
{${\bar K}\Xi_c^\prime$}   &     & $1$ & $\sqrt{1/2}~\kappa_c$ & $-\sqrt{6}$ & $\m\m 0$   \\
{ $D\Xi$} &     &     &    $\m 2$        &  $-\sqrt{1/3}~\kappa_c$  & $-\sqrt{2/3}~\kappa_c$ \\
{ $\eta\Omega_c^0$} &     &     &     &  $\m\m 0$  &  $\m\m 0$\\
{ $\eta^\prime\Omega_c^0$}  &     &     &    &     &  $\m\m 0$   \\             
\hline \\
\end{tabular}
\end{center}
\label{tab:coeff}
\vspace{-1.0cm}
\end{table}

In the $VB$ case, the allowed states are $D^*\Xi (3326)$, $\bar{K}^*\Xi_c (3363)$, $\bar{K}^*\Xi'_c (3470)$, $\omega \Omega_c (3480)$, $\phi \Omega_c (3717)$, $\bar{D}_s^* \Omega_{cc} (5662)$ and $J/\psi \Omega_c (5794)$, where, again, we neglect the doubly charmed states. The coefficients $C_{ij}$ can be straightforwardly obtained from those in Table~\ref{tab:coeff} with: $\pi \rightarrow \rho,\, K\rightarrow K^\ast,\, \bar{K}\rightarrow\bar{K}^*,\,  D\rightarrow D^*,\, \bar{D}\rightarrow\bar{D}^*,\, {1/\sqrt{3}}\eta+\sqrt{2/3}\eta'\rightarrow\omega$ and $-\sqrt{2/3}\eta+{1/\sqrt{3}}\eta'\rightarrow\phi$.

Resonance poles of the scattering amplitude appear in the {\it second Riemann sheet} of the complex energy plane. The residues at the pole position $z_p$ give the coupling constants $g_i$ of the resonance to the various channels and the real part of $-g_i^2(\partial G/\partial\sqrt{s})|_{z_p}$ corresponds to the compositeness, i.e., the amount of $i^{\rm th}$-channel meson--baryon component.

\section{Results}
The values of the subtraction constant, $a_l(\mu=1\rm~GeV)$, used when solving Eq.~(\ref{eq:BSeq}) are chosen so as the loop function in dimensional regularization coincides  with the loop function regularized with a cut-off $\Lambda=800\rm\,MeV$ at the channel threshold (``Model 1"). The resulting $PB$ scattering amplitude shows two poles,
\begin{eqnarray}
M_1 =  {\rm Re}z_1= 3051.6\,{\rm MeV},&\quad\,\,\Gamma_1 = -2 {\rm Im}z_1= 0.45\,{\rm MeV} \nonumber \\
M_2 =  {\rm Re}z_2 = 3103.3\,{\rm MeV},&\quad\Gamma_2 = -2 {\rm Im}z_2= 17\,{\rm MeV\,,}
\label{eq:reson_800}
\end{eqnarray}
corresponding to resonances with spin--parity $J^P=1/2^-$. Their energies are very similar to the second and fourth $\Omega_c^0$ states discovered by LHCb \cite{Aaij:2017nav}.
% , with properties:
% \begin{eqnarray}
% \Omega_c(3050)^0:~~& M=3050.2\pm0.1\pm0.1^{+0.3}_{-0.5}~{\rm MeV}, \nonumber \\
%                                  &\Gamma=0.8\pm0.2\pm0.1~{\rm MeV},\nonumber \\
% \Omega_c(3090)^0:~~& M=3090.2\pm0.3\pm0.5^{+0.3}_{-0.5}~{\rm MeV}, \nonumber \\
%                                  &\Gamma=8.7\pm1.0\pm0.8~{\rm MeV}.
% \label{eq:exp}
%  \end{eqnarray}

These results clearly show that the meson--baryon dynamical models are able to generate states in the energy range of interest, although the mass of our heavier state is larger by 10\,MeV and its width is about twice the experimental one. In an attempt to explore the possibilities of our model, we let the values of the five subtraction constants vary freely within a reasonably constrained range and look for a combination that reproduces the characteristics of the two observed states, $\Omega_c(3050)^0$
and $\Omega_c(3090)^0$, within $2\sigma$ of the experimental errors. 
Table~\ref{tab:pseudo} displays the new properties of the poles for a representative set of $a_l(\mu=1\rm~GeV)$ with equivalent cut-off values in the $320$--$950\;\rm MeV$ range (referred as ``Model 2" in \cite{Montana:2017kjw}). We note that the strongest change corresponds to $a_{\bar{K}\Xi_c}$, needed to decrease the width of the $\Omega_c(3090)^0$. Its equivalent cut-off value of 320 MeV is on the low side of the usually employed values but it is still naturally sized.

\begin{table}[hbt!]
%\vspace{-0.3cm}
\caption{Position ($\sqrt{s}=M-\I\Gamma/2$), couplings and compositeness of the $\Omega^0_c$ states generated employing ``Model 2".}
\begin{center}
\begin{tabular}{lcccc}
%\hline \\[-2.5mm]
\hline 
&    \multicolumn{2}{c}{}    &   \multicolumn{2}{c}{} \\   [-3mm]
\multicolumn{5}{c}{ {\bf $0^- \otimes 1/2^+$} interaction in the {\bf$(I,S,C)=(0,-2,1)$} sector } \\
\hline
&    \multicolumn{2}{c}{}    &   \multicolumn{2}{c}{} \\   [-3mm]
$M\;\rm[MeV]$             &    \multicolumn{2}{c}{$\qquad 3050.3$}    &    \multicolumn{2}{c}{$\qquad 3090.8$}   \\
$\Gamma\;\rm[MeV]$   &     \multicolumn{2}{c}{$\qquad 0.44$}        &    \multicolumn{2}{c}{$\qquad 12$}     \\ 
\hline
 &    \multicolumn{2}{c}{}    &   \multicolumn{2}{c}{} \\  [-3mm]
 &      $\qquad | g_i|$    & $-g_i^2 dG/dE$    &   $\qquad | g_i|$  & $-g_i^2 dG/dE$ \\
$\bar{K}\Xi_c (2964)$   &  $\qquad 0.11$  & $\m 0.00+\I\,0.00$     &    $\qquad 0.49$  & $-0.02+\I\,0.01$  \\
$\bar{K}\Xi'_c (3070)$  &  $\qquad 1.80$  & $\m 0.61+\I\,0.01$     &    $\qquad 0.35$  & $\m 0.02-\I\,0.02$  \\
$D\Xi (3189)$           &  $\qquad 1.36$  & $\m 0.07-\I\,0.01$     &    $\qquad 4.28$  & $\m 0.91-\I\,0.01$ \\
$\eta \Omega_c (3246)$  &  $\qquad 1.63$  & $\m 0.14+\I\,0.00$     &    $\qquad 0.39$  & $\m 0.01+\I\,0.01$ \\
$\eta' \Omega_c (3656)$ &  $\qquad 0.06$  & $\m 0.00+\I\,0.00$     &    $\qquad 0.28$  & $\m 0.00+\I\,0.00$ \\
\hline
\end{tabular}
\end{center}
\label{tab:pseudo}
\vspace{-0.7cm}
\end{table}

We also show in table~\ref{tab:pseudo} the couplings of each resonance to the various meson--baryon channels and the corresponding compositeness. The lowest energy state at 3050\,MeV has a strong coupling to $\bar{K}\Xi'_c$ and has a high compositeness in this channel but also couples appreciably to $D\Xi $ and $\eta \Omega_c$ channels. The higher energy resonance at 3090\,MeV couples strongly to $D\Xi$ and clearly qualifies as a $D\Xi $ bound state with a compositeness in this channel of 0.91.

\begin{figure}[htb]
\vspace{-0.5cm}
\centering
  \includegraphics[width=0.45\textwidth]{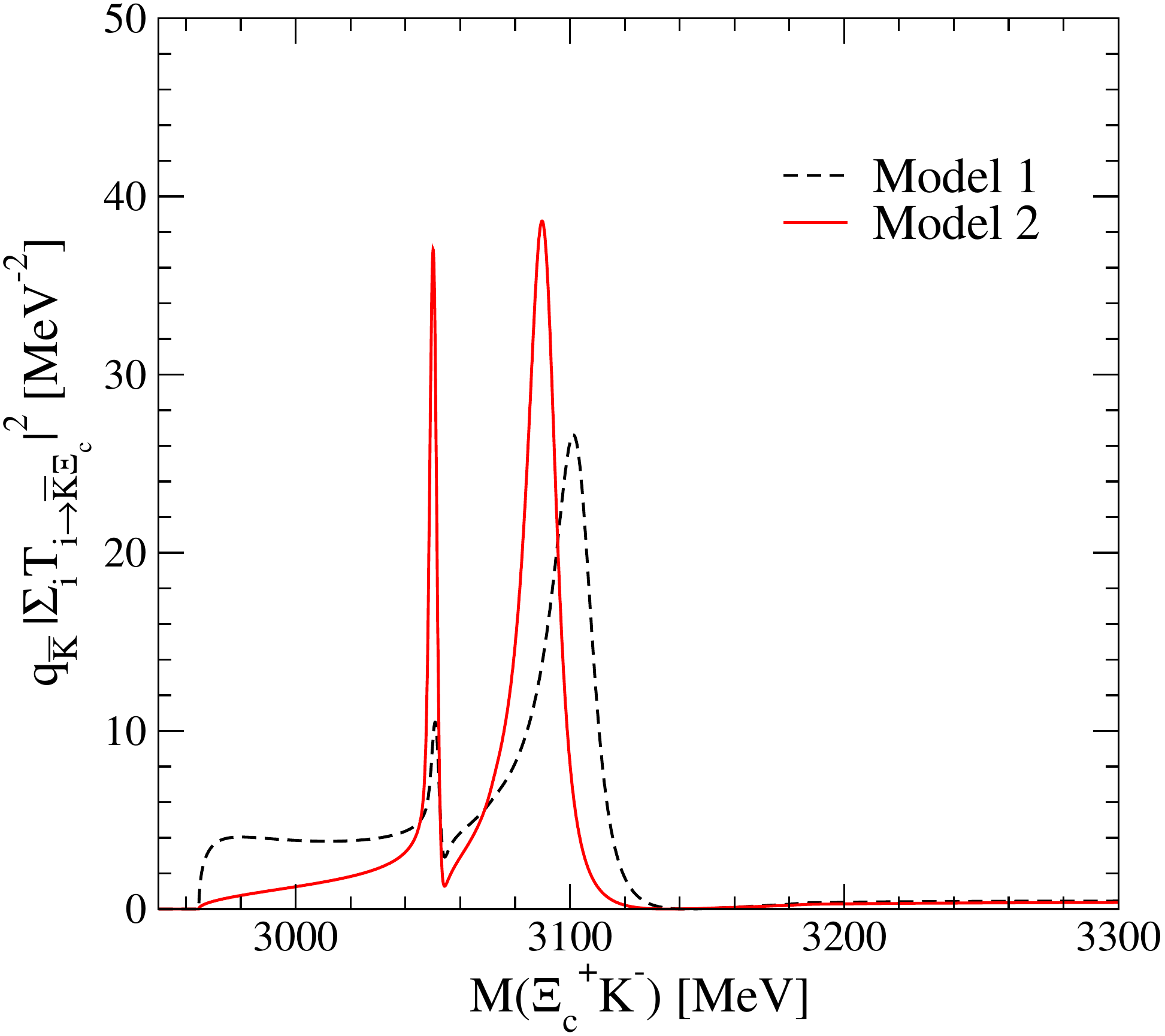}\hspace{0.5cm}%%
%\vspace{0.5cm}
\begin{minipage}[b]{0.5\textwidth}\caption{(Colour online) Sum of amplitudes squared times the momentum of the $K^-$ versus the $\bar{K}\Xi_c$ energy in the centre-of-mass frame, where $T_{i\to \bar{K}\Xi_c}$ is the amplitude for the $i \to \bar{K}\Xi_c$ transition obtained here with either ``Model 1" (black dashed line) or ``Model 2" (red solid line), with $i$ being any of the five coupled channels. The $q_{K^-}$ acts as a phase-space modulator. The calculated spectrum has been convoluted with the energy dependent resolution of the experiment.} \vspace{0.2cm}\label{fig:t2}
\end{minipage}
\vspace{-0.5cm} 
\end{figure}

The five $\Omega_c^0$ states were observed from the $K^-\Xi^+_c$ invariant mass spectrum obtained in high energy $pp$ collision data by the LHCb \cite{Aaij:2017nav}, which is tremendously difficult to model. In Fig.~\ref{fig:t2} we display a merely illustrative plot of the spectrum that our models would predict that retains certain similarities with the spectrum of Fig.~2 in Ref.~\cite{Aaij:2017nav} in the energy regions of the 3050\,MeV and 3090\,MeV states.

\begin{figure}[h]
\vspace{-0.5cm}
\includegraphics[width=0.57\textwidth]{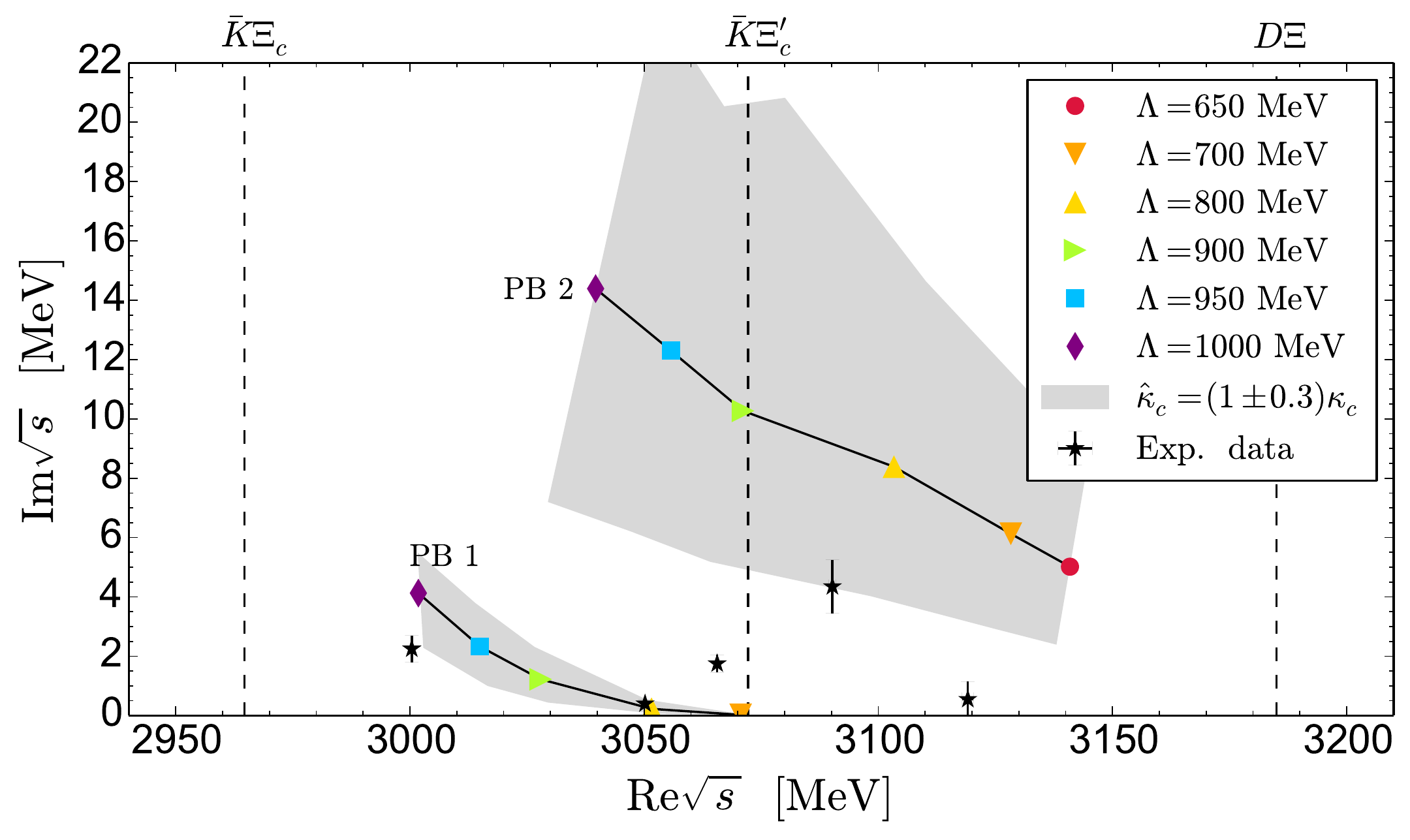}\hspace{0.5cm}%%
\begin{minipage}[b]{0.38\textwidth}\vspace{0.2cm} \caption{Evolution of the position of the resonance poles for various cut-off values. The grey area indicates the region of results covered when a variation of 30\% in the $SU(4)$ breaking is assumed in the transitions mediated by heavy-meson exchange.}\vspace{0.5cm}\label{fig:cut-off}
\end{minipage}
\vspace{-0.5cm}
\end{figure}

Next we discuss the dependence of these results on the assumed value of the cut-off, as well as the influence of a certain amount of $SU(4)$ symmetry violation associated to the fact that the charm quark is substantially heavier than the light quarks. The solid lines in Fig.~\ref{fig:cut-off} indicate the evolution of the poles as the value of the cut-off is increased from 650\,MeV to 1000\,MeV. On the other hand, we note that the violation of $SU(4)$ symmetry is already partly implemented by the use of the physical meson and baryon masses in the interaction kernel. Moreover, only $SU(3)$ is effectively acting in the transitions mediated by light vector mesons and thus these will be left untouched. Therefore, up to an additional $30\%$ of $SU(4)$ breaking is implemented only in the matrix elements that connect states via the t-channel exchange of a charmed vector meson, and this is achieved by allowing the factor $\kappa_c$ to vary in the range $(0.7-1.3)\kappa_c$. The grey area in Fig.~\ref{fig:cut-off} corresponds to the region in the complex plane where the resonances can be found varying both the cut-off and the amount of $SU(4)$ violation. The fact that these band of uncertainties includes the 3050\,MeV and the 3090\,MeV resonances measured at LHCb reinforces their interpretation as meson--baryon molecules.

In the case of $VB$ scattering we have followed a similar procedure to look for resonances, which are degenerate in spin, $J^P=1/2^-,3/2^-$. Employing subtraction constants mapped onto a cut-off of $\Lambda=800$\,MeV, we see a similar pattern as that found for the $PB$ case. A lower energy resonance mainly classifying as a $D^*\Xi$ molecule appears at 3231\,MeV and a higher energy resonance is generated at 3419\,MeV and corresponds to a $\bar{K}^*\Xi'_c $ composite state with some admixture of $\omega \Omega_c^0$ and $\phi \Omega_c$ components.  There is an additional pole in between these two, coupling strongly to $\bar{K}^*\Xi_c$ states. These three resonances are located in an energy region above the states reported by the LHCb collaboration where no narrow structures have been seen \cite{Aaij:2017nav}. We note, however, that the states found here from the $VB$ interaction are artificially narrow as they do not couple to, and hence cannot decay into, the $PB$ states that lie at lower energy.

Finally, we show the results of extending our model to the bottom sector by employing the meson--baryon interaction kernels obtained from the Lagrangians of Eqs.~(\ref{eq:vertexVPP})--(\ref{eq:vertexVVV}), but replacing the charm mesons and baryons by their bottom counterparts (see the details in \cite{Montana:2017kjw}). A coefficient $\kappa_b = 0.1$ in certain non-diagonal transitions that accounts for the much larger mass of the exchanged bottom vector mesons with respect to the light ones is the analogous to $\kappa_c$.

Our results for the $\Omega_b^-$ resonances are very similar to those found in the charm sector. The $PB$ interaction generates two states at 6418\,MeV and 6519\,MeV with spin $J^P=1/2^-$, the former couples strongly to ${\bar K}\Xi^\prime_c$ and $\eta\Omega_b$ while the later is essentially a $B\Xi$ bound state. In the $VB$ interaction we find $J=1/2^-,3/2^-$ spin degenerate $\Omega^-_b$ states at 6560\,MeV, coupling strongly to ${\bar B}^*\Xi$, 6665\,MeV, coupling  to $K^*\Xi_b$, and 6797\,MeV, being a mixture of $\omega \Omega_b$, ${\bar K}^* \Xi'_b$ and $\Phi \Omega_b$.

\section{Conclusions}
Employing a t-channel vector meson exchange model with effective Lagrangians, we have studied the interaction of the low-lying pseudoscalar and vector mesons with the ground-state baryons in the charm $+1$, strangeness $-2$ and isospin $0$ sector. Two resonances with energies and widths very similar to some of the $\Omega_c^0$ states discovered recently at LHCb are found in the unitarized scattering amplitudes of the interaction of pseudoscalar mesons with baryons. We have extended the model to the bottom sector and predicted several $\Omega_b^-$ resonances in the energy region $6400$--$6800$\,MeV with a molecular meson--baryon structure.

Several other works \cite{Debastiani:2017ewu,Wang:2017smo,Chen:2017xat,Nieves:2017jjx} have also addressed the possibility of interpreting of some of the $\Omega_c$ states seen at CERN as quasi-bound meson--baryon systems, as well as the prediction of analogous states in the bottom sector \cite{Liang:2017ejq}, finding results which are similar to those of our work \cite{Montana:2017kjw} and hinting that the meson--baryon description cannot be ignored when trying to understand the nature of these excited heavy baryons.

\section*{Acknowledgments}
This work is supported by the Spanish MINECO under the contracts MDM-2014-0369, FIS2017-87534-P and the doctoral grant FPU17/04910 and by the Generalitat de Catalunya under the doctoral grant 2018 FI\_B 00234.

\bibliography{references}

\end{document}